\documentclass[preprint,12pt]{elsarticle}

\usepackage{amssymb}
\usepackage{amsmath}
\usepackage{amsthm}
\usepackage{graphicx}
\usepackage{natbib}
\usepackage{hyperref}
\usepackage{caption}
\journal{Cyber Security and Applications}

\begin{document}

\begin{frontmatter}

\title{Detecting Fileless Cryptojacking in PowerShell Using AST-Enhanced CodeBERT Models}

\author[uc-it]{Said Varlioglu}
\ead{varlioms@mail.uc.edu}

\author[uc-it]{Nelly Elsayed\corref{cor1}}
\ead{elsayeny@ucmail.uc.edu}

\author[uc-it]{Murat Ozer}
\ead{ozermm@ucmail.uc.edu}

\author[uc-it]{Zag ElSayed}
\ead{elsayezs@ucmail.uc.edu}

\author[uc-it,uc-ece]{John M. Emmert}
\ead{emmertj@ucmail.uc.edu}

\cortext[cor1]{Corresponding author}
%\ead{ozermm@ucmail.uc.edu}
%\ead{elsayezs@ucmail.uc.edu}
%\ead{emmertj@ucmail.uc.edu}

%% Abstract
\begin{abstract}
With the emergence of remote code execution (RCE) vulnerabilities in ubiquitous libraries and advanced social engineering techniques, threat actors have started conducting widespread fileless cryptojacking attacks. These attacks have become effective with stealthy techniques based on PowerShell-based exploitation in Windows OS environments. Even if attacks are detected and malicious scripts removed, processes may remain operational on victim endpoints, creating a significant challenge for detection mechanisms. In this paper, we conducted an experimental study with a collected dataset on detecting PowerShell-based fileless cryptojacking scripts. The results showed that Abstract Syntax Tree (AST)-based fine-tuned CodeBERT achieved a high recall rate, proving the importance of the use of AST integration and fine-tuned pre-trained models for programming language.
\end{abstract}


%item Provides the first comprehensive socio-technical framework for governing AI-driven clinical speech-to-text documentation systems.
%\item Highlights accuracy disparities for patients with speech disorders, accented speech, or neurological conditions, raising equity concerns.
%\item Offers actionable governance strategies for technical performance validation, clinician training, patient consent, and data protection.
%\item Identifies critical transparency, reliability, and privacy risks across technical, human, ethical, organizational, and regulatory layers.
%\item Demonstrates that effective adoption of STT systems requires integrated oversight, not solely improvements in model accuracy.

%% Keywords
\begin{keyword}
Fileless Malware \sep Cryptojacking \sep  PowerShell \sep  Machine Learning \sep  Cybersecurity



\end{keyword}

\end{frontmatter}



\section{Introduction}

Cryptojacking, an unauthorized cryptocurrency mining on infected devices, is a popular cyber attack based on financial motivations \cite{CryptojackingTrap,BotcoinTrap}. Detection systems face challenges because of fileless attack techniques. Fileless attacks have become popular with the prevalence of ransomware and cryptojacking since 2017 \cite{olaimat2021ransomware,hendler2018detecting,varlioglu2022dangerous}.

In a fileless attack, there is no file to inspect, but commands and network activity are detectable. Fileless cryptojacking resides in memory (RAM) without touching the disk on a computer. In Windows systems, PowerShell plays a significant role in these advanced volatile attacks with its exploitable features \cite{bulazel2017survey,varlioglu2022dangerous}.

PowerShell-based fileless cryptojacking attacks can be initiated with phishing emails, zero-day vulnerability exploitations, or hidden scripts on malicious websites \cite{trendmicro2019b}. Malicious scripts are often stored and downloaded from text storage web services, such as Pastebin \cite{Gundert}, as observed in fileless ransomware attacks \cite{moussaileb2021survey} \cite{sophos2019c}. 

PowerShell can access critical Windows system functions with remote content retrieval, in-memory command executions, and access to local registry keys and scheduled tasks \cite{hendler2018detecting}. Malware authors exploit it by interacting with the operating system and almost all Microsoft software, from the Office suite to the SQL Server database engine. \cite{rousseau2017hijacking,xu2021research}.

Its adaptability helps threat actors reduce the need to customize payloads or download overtly malicious tools to infected systems. They can filelessly load and execute commands using its invocation feature without touching the disk or creating new system processes \cite{RedCanary2023}. 

With this fileless nature, even if attacks are detected, and the original malicious scripts are identified and removed, the processes may remain operational at the victim endpoints \cite{varlioglu2022dangerous}.

In some cases, fileless malware may exhibit indirect file activity while establishing a mining service or persistence mechanism. However, the initial phases of the attacks are still fileless \cite{Microsoft82021}.

\begin{figure*}
    \centering
    \includegraphics[width=0.6\textwidth, , alt={An example of the encoded PowerShell-based fileless cryptojacking script code screenshot.}]{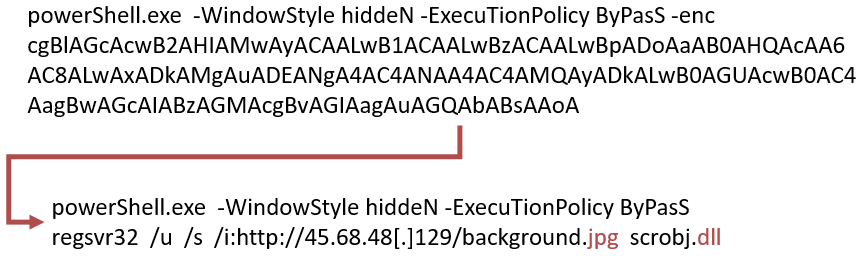}
    \caption{Sample encoded PowerShell-based fileless cryptojacking script.}
    \label{sample_ps_cryptomining_script}
\end{figure*}

\begin{figure}
    \centering
    \includegraphics[width=0.45\textwidth,alt={A Sample PowerShell-based fileless cryptojacking attacks workflow. The flow consists of the 10 stages.}]{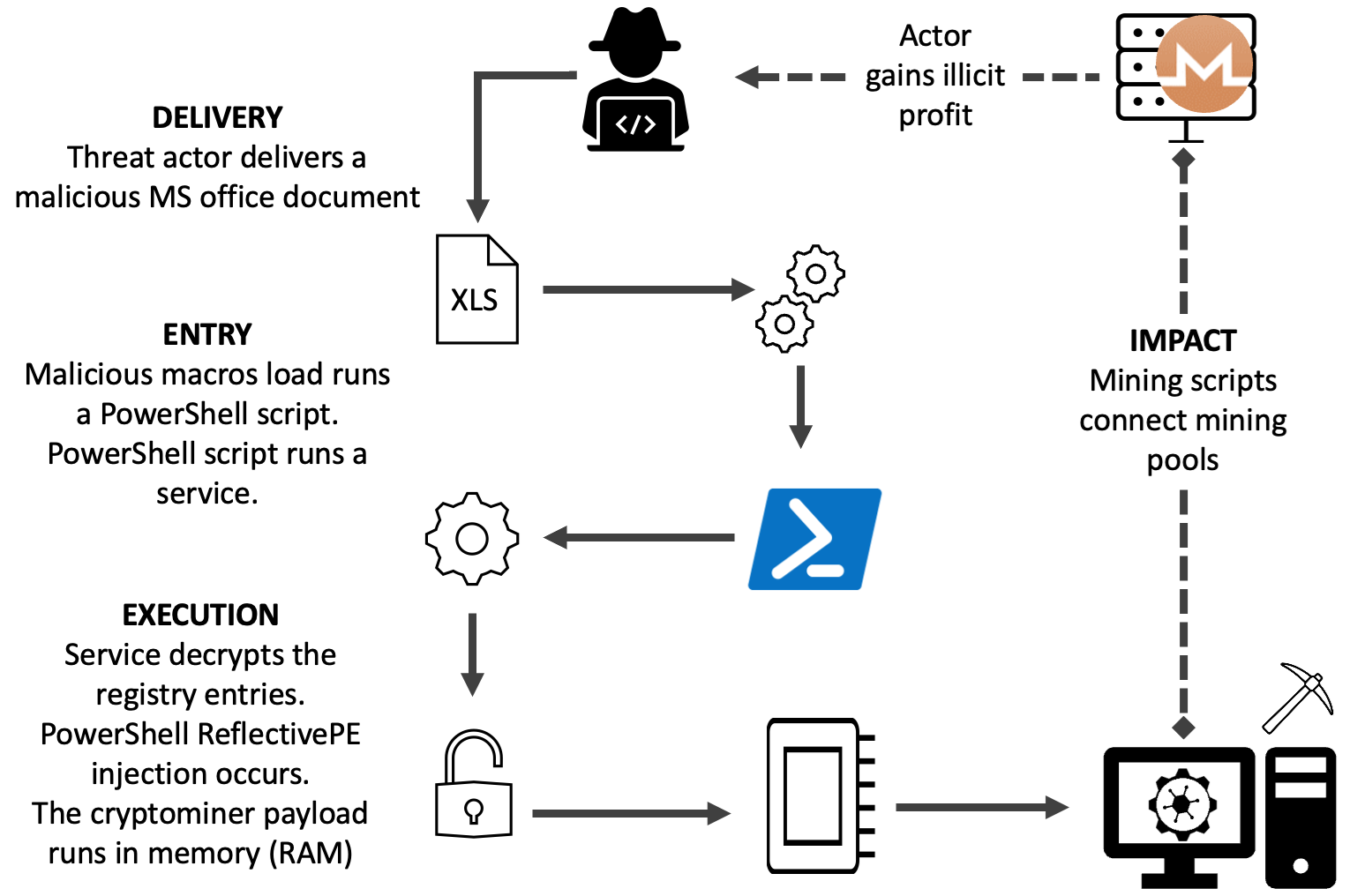}
    \caption{Sample PowerShell-based fileless cryptojacking attacks workflow.}
    \label{psbased}
\end{figure}

As seen in Fig.~\ref{sample_ps_cryptomining_script} and Fig.~\ref{psbased}, threat actors deliver malicious Excel documents that contain malicious office macros to deploy a DLL downloader via the register server (regsvr32). The macros trigger malicious encoded PowerShell scripts to conduct process injection and add new registry entry. Finally, the CobaltStrike payload, an exploitable penetration testing script, is injected into a legitimate process to establish C2. The encoded payload is decoded and run on memory to deploy cryptomining \cite{Unit42_1,Mandiant1,Newsroom_2024}.

In fileless attacks such as those involving Cobalt Strike, scripts have been observed employing obfuscation techniques such as Base64-encoded commands as seen in Figure~\ref{psbased} with GunZip compression and encrypted XOR keys \cite{varlioglu2024pulse}. These methods hide specific Windows API calls made by the shellcode and process injection into legitimate processes, and they also mask the establishment of Command and Control (C2) connections and secure session data with encryption \cite{Newton2020}.

With the emergence of new Remote Code Execution (RCE) vulnerabilities, such as Log4Shell (CVE-2021-44228) and SolarWinds Orion (CVE-2020-10177), which affect numerous software products across various industries, widespread fileless cryptojacking and ransomware attacks have become prevalent \cite{SzappanosGallagher2022}. Additionally, threat actors use various open-source frameworks, including PowerShell Empire and PowerSploit, which are designed to support PowerShell scripting in post-exploitation cyber-offensive activities \cite{Mandiant1,hendler2018detecting}.

PowerShell-based fileless cryptojacking attacks can involve the creation of encrypted or decoded registries containing decryption routines, using reflective Portable Executable (PE) loaders for executing malicious code, and deploying process injection. These are key features of fileless attacks that provide sophisticated evasion techniques and persistence mechanisms \cite{Ang_Gelera_Villanueva_2018}. In particular, a reflective PE loader is commonly used for process injection. This technique allows the malware to load and execute its payload directly into memory, bypassing traditional file-based detection mechanisms \cite{varlioglu2024pulse}.

Since fileless threats can provide attackers with command and control abilities using backdoors \cite{moussaileb2021survey}, a fileless cryptojacking attack can be converted into a data exfiltration activity.

In the literature, there are various kinds of research on specifically in-browser and in-host cryptojackings with the detection of their attack patterns, especially in network behavior \cite{Hong2018,ParraRodriguez2018,sachan2022dns,Saad2018,CryptojackingTrap,BotcoinTrap,khan2023real,razali2019cmblock,naseem2021minos}. Also, there is various research to identify malicious PowerShell scripts, in general, using machine learning techniques \cite{luheuristic, alahmadi2022mpsautodetect,song2021evaluations,mimura2021static,hendler2018detecting,rubin2019amsi,hung2024machine,rusak2018ast,fang2021effective,khalid2023insight,liguori2024power}. However, there is a gap in that no research specifically addressed PowerShell-based fileless cryptojacking. 

Hence, there is a crucial need for  research in this domain. To our best knowledge, no research has been conducted on detecting PowerShell-based fileless cryptojacking using machine learning. This gap can be filled out with a unique dedicated dataset derived from those attacks.

We attempted to fill this research gap by collecting a unique dataset and applying machine learning models with an experimental study on detecting PowerShell-based fileless cryptojacking scripts. The results showed that Abstract Syntax Tree (AST)-based fine-tuned CodeBERT achieved a high recall rate, proving the importance of using AST integration and fine-tuned pre-trained models for programming language.

\section{Method} 

\label{ML_Method}

The research body on fileless malware detection have used machine learning models to identify malicious PowerShell scripts \cite{luheuristic,alahmadi2022mpsautodetect,song2021evaluations,mimura2021static,hendler2018detecting,rubin2019amsi,hung2024machine,rusak2018ast,fang2021effective,khalid2023insight,liguori2024power}. Researchers commonly used deep learning techniques with natural language processing approaches \cite{rusak2018ast,yang2023powerdetector}. However, these solutions try to detect general malicious scripts. There is a need in the literature that machine learning models can also be used to improve detections on specifically PowerShell-based fileless cryptojacking attacks.

In this paper, we use a new mixed method for code representations of PowerShell scripts with their Abstract Syntax Trees (ASTs) using the selected machine learning models, Long Short-Term Memory (LSTM), Bidirectional LSTM (BiLSTM), and CodeBERT. The goal is to prepare this mixed approach for model training to detect cryptojacking scripts with other general PowerShell scripts. The assumption is that the mixed method provides effective detection mechanisms for detecting cryptojacking scripts as well as other general malicious PowerShell scripts.

Furthermore, the assumption is that developing light and fast-based detection models instead of relying on manually designed features and specific feature extraction can help the detections rather than complex approaches such as feature extractions and de-obfuscation processes.

First, we conduct a background literature review on the existing and latest solutions to prepare an experimental base for machine learning models. This phase examines Abstract Syntax Tree (AST) representations of PowerShell scripts that can be useful to prepare data inputs for training in the selected machine learning models. We also reviewed existing models for this domain to prepare and select the appropriate approach for machine learning models. 

Second, we collected a dataset that contains 500 collected malicious PowerShell based fileless cryptojacking scripts. Those scripts include Purple Fox, Lemon Duck, and Tor2Mine malware families. We merged our dataset with a secondary dataset \cite{fang2021effective} that contains 1,770 malicious scripts and 4,819 benign scripts but does not specifically target cryptojacking. Because cryptojacking also embeds general malicious PowerShell scripts to carry out attacks. We used those datasets to train and evaluate machine learning models.

In addition, we followed the cleaning and pre-processing steps as proper inputs for machine learning algorithms. The selected machine learning models were trained and evaluated.

The results were analyzed to determine the effectiveness of each model in detecting fileless cryptojacking scripts. We evaluate both AST-based and non-AST-based approaches to see the effectiveness of this mixed approach. Instead of heavily modifying models, focusing on feature extraction, or undergoing the complex process of obfuscation-deobfuscation examinations with an AST-based approach provides flexibility. This flexible approach is powered by an AST-pipeline-based examination with minimally modified versions of common machine learning models.

Since it is hard to find malicious samples and easy to find benign samples, the number of benign samples is higher. However, the ratio is not heavily skewed. This helps ensure that the machine learning models receive a fair representation of both categories, reducing bias. 

The comparison of file sizes shows an overlapping distribution, and malicious scripts have a slightly lower median file size compared to benign ones. Some malicious scripts are significantly larger due to obfuscation or all-in-one script features. Both benign and malicious scripts have similar lengths with a concentration below 500 lines. Malicious scripts have a slightly wider spread due to longer scripts. Malicious script's entropy is higher due to more obfuscation, encoding, "for loops," and "while loops" in the longer scripts. This also shows a pattern that benign scripts commonly contain more readable and structured content than malicious scripts.

\section{Background Literature}

In the following, we conduct a review of the background literature on the existing and latest solutions to prepare an experimental base for machine learning models. This phase examines Abstract Syntax Tree (AST) representations of PowerShell scripts that can be useful to prepare data inputs for training in the selected machine learning models. We also reviewed existing models for this domain to prepare and select the appropriate approach for machine learning models.

\subsection{Abstract Syntax Tree (AST)}

Abstract Syntax Tree (AST), a tree-like version of the code to show the abstract syntactic structure of the source code \cite{crew1997astlog}, can be used for malicious script detection.  It is used in both programming languages and software development tools \cite{zhang2019novel} to clearly explain code analysis \cite{holz2019investigating}.

Specifically, AST breaks down the code into its components and organizes them hierarchically as a tree-style version. Each node in the tree represents an element of the language with the expressions, statements, or operators \cite{rusak2018ast}. By analyzing the AST, we can understand the main logic of the code without looking into the specific syntax or formatting details. This allows for a more precise evaluation of the code's potential impacts and risks \cite{COOPER2023159,holz2019investigating}.

AST analysis is useful for identifying patterns, detecting potential vulnerabilities, and ensuring compliance with coding standards \cite{rusak2018ast}. It provides information on the functionality of the code and potential risks as effective code review and security assessment processes \cite{COOPER2023159,holz2019investigating}.

\begin{figure}
    \centering
    \vspace{0.1cm}
    \includegraphics[width=0.45\linewidth, alt={ping command ping -c 4 -t 64 uc.edu}]{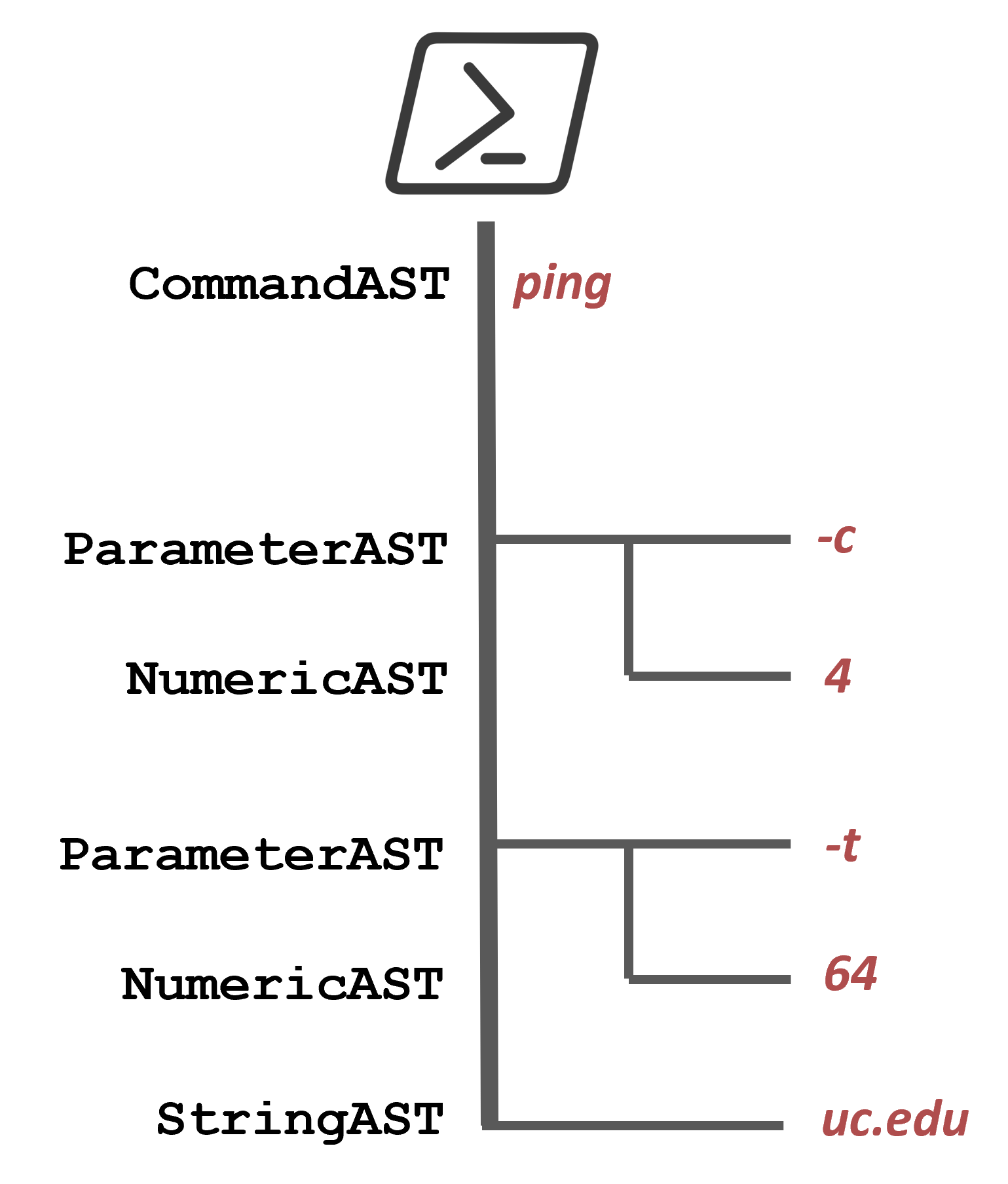}
    \caption{Sample AST of a Ping command: “ping -c 4 -t 64 uc.edu".}
    \label{ast_ping}
\end{figure}

As shown in Fig. \ref{ast_ping}, it breaks down a ping command into hierarchical components. This feature can help machine learning models to learn the patterns with the hierarchical structure and sequential data.

\subsection{Abstract Syntax Tree (AST) Types}

In order to show the types of AST, the sample PowerShell command is showed to break down. This tree structure shows the root and child nodes based on the AST types.

The sample command below is a Purple Fox cryptojacking malicious command for a C2 connection. The PowerShell script downloads a backdoor written in .NET implanted during the intrusion. The backdoor is FoxSocket, which is a Purple Fox-related malware that leverages WebSockets to communicate with its command-and-control (C2) servers. The command downloads a malicious payload from the specified URLs with an MSI command as an argument to install the MSI package as an admin without any user interaction \cite{tremdmicropurplefox2021}.

\begin{verbatim}
	powershell.exe -nop -exec bypass -c
	IEX (New-Object Net.WebClient).DownloadString(
	'http[:]//117.187.136.141[:]13405/57BC9B7E.Png');
	MsiMake http[:]//117.187.136.141[:]13405/0CFA042F.Png
\end{verbatim}

The AST types of the sample command above \cite{AST_Types}:

\begin{itemize}
    \item \textbf{Expression} 
    \begin{itemize}
        \item {CommandExpressionAst}: Represents the expression.
        \item {CommandInvocationAst}: Invokes the parameters.
        \item {ExpressionAst}: Represents the expression invoked.
        \item {TypeNameAst}: Net.WebClient
        \item {MethodInvocationAst}: Method invocation.
        \item {MethodNameAst}: DownloadString
        \item {StringLiteralAst}: '<URL>:13405/57BC9B7E.Png'
        \item {CmdletAst}: MsiMake
     \end{itemize}
\end{itemize}

\begin{itemize}
    \item \textbf{Statement}
\begin{itemize}
    \item {PipelineAst}: Pipeline of commands executed.
    \item {CommandInvocationAst}: Execute Invoke-Expression.
    \item {ExpressionAst}: Represents IEX.
    \item {MethodInvocationAst}: Method call DownloadString.
    \item {ArgumentAst}: URL string literal.
    \item {CmdletAst}: MsiMake command with its argument.
\end{itemize}
\end{itemize}

The Abstract Syntax Tree (AST) tree format:

\begin{verbatim}
__________________________________________
PowerShell Cmdlet
Argument: -nop
Argument: -exec bypass
Argument: -c
|   IEX (New-Object Net.WebClient).
DownloadString('PAYLOAD')
|   Invoke-Expression (IEX)
|   |   MethodCall
|   |   |   New-Object (Net.WebClient)
|   |   |   |   TypeName: Net.WebClient
|   |   |   Method: DownloadString
|   |   |   |   Argument
|   |   |   |   |   StringLiteral: PAYLOAD
|   |   Cmdlet or Function Call: MsiMake
|   |   |   Argument
|   |   |   |   StringLiteral: PAYLOAD
__________________________________________
\end{verbatim}

In order to create the pipeline of ASTs, each PowerShell script is converted into ASTs using built-in parser class:

\noindent
[System.Management.Automation.Language.Parser] of PowerShell \cite{schubert2024malware}. 

In AST-based machine learning models, the pipeline creation process includes AST report generation for each malicious and benign script and tokenizing them \cite{rusak2018ast, fang2021effective, nurmi2019script, rose2024scriptblock, xu2021research, miao2023ast2vec}. Also, AST (Abstract Syntax Tree) pipelines are being examined in a safe environment to handle malicious PowerShell scripts safely.

\subsection{Long Short-term Memory (LSTM)}

The long short-term memory (LSTM) network is a type of recurrent neural network (RNN), a type of artificial neural network that processes sequential data, like text \cite{elsayed2019deep}. LSTM is used to predict temporal-dependent data such as time series \cite{elsayed2019gated,azumah2021deep}. It is outperforms the RNN because it solves the problem of vanishing gradients \cite{elsayed2020reduced, hochreiter1997long}.

The LSTM networks were proposed by Hochreiter et al. \cite{hochreiter1997long} in 1997 and became popular with the successful results on the models of natural language processing (NLP) with better sentiment analysis and classification. LSTM approach is capable of capturing high-level abstractions and reducing dimensionality \cite{alshemali2020improving,gimenez2020semantic,guo2020gluoncv,park2020deep, abualigah2020sentiment,balamurali2020develop, fischer2018deep,elsayed2022litelstm, unanue2017recurrent,luo2017recurrent, jelodar2020deep}.  

An LSTM unit consists of a cell, an input gate, an output gate, and a forget gate \cite{gers2000learning}. The cell retains information over varying time intervals, while the three gates control the flow of information. It is a designed model for classifying sequential data with long-term dependencies. It improved predictions in text classification \cite{hochreiter1996lstm}, making it useful for classifying malicious and benign scripts.

\begin{figure}
    \centering
    \includegraphics[width=8cm, height = 6cm, alt={The LSTM block architecture diagrams including the three gates and the activation functions.}]{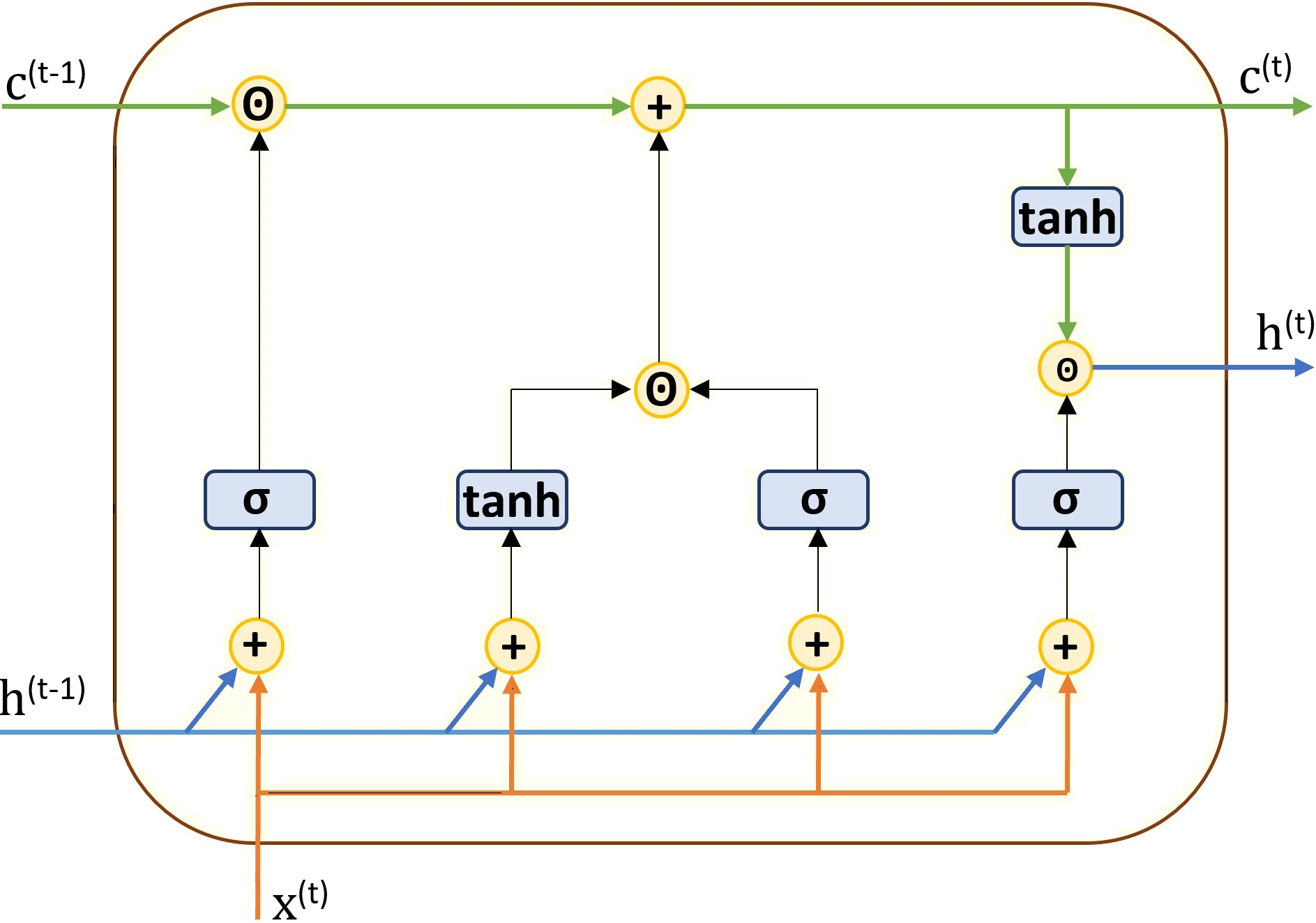}
    \caption{The LSTM architecture diagram~\cite{elsayed2023litelstm}.}
    \label{lstm_block1}
\end{figure}

Fig.~\ref{lstm_block1} shows the LSTM architecture components where each memory cell state (c) in an LSTM network serves a specific function \cite{azumah2021deep} to capture the long-term and short-term dependencies in the data. Input gate (i) decides new information from the input \textit{x} for the cell state memory (c). Output gate (o) specifies the output from the memory cell state. Forget gate (f) decides what to discard from the memory cell state.

Input gates determine which new information to store in the current state, and output gates assign a value from \textit{0} to \textit{1} to decide which information from the current state to output. This selective output of relevant information helps LSTM networks maintain important long-term dependencies for making predictions in both current and future time steps \cite{elsayed2019gated, azumah2021deep}. Forget gates determine which information from the previous state is discarded by assigning a value between \textit{0} and \textit{1}, with \textit{1} indicating that the information will be retained and zero indicating that it will be discarded \cite{elsayed2019gated, azumah2021deep}.  The output \textit{h} can be calculated by:
\begin{align}
	i^{(t)}&= \sigma(W_{xi} x^{(t)} +U_{hi}  h^{(t-1)}+ b_i)\label{eqn:i_lstm}\\ 
	g^{(t)}&= \mathrm{tanh}(W_{xg} x^{(t)} +U_{hg}  h^{(t-1)}+ b_g)\label{eqn:g_lstm}\\ 
	f^{(t)}&=\sigma(W_{xf} x^{(t)} +U_{hf}  h^{(t-1)} + b_f)\label{eqn:f_lstm}\\ 
	o^{(t)}&=\sigma(W_{xo} x^{(t)} +U_{ho}  h^{(t-1)} + b_o)\label{eqn:o_lstm}\\ 
	c^{(t)}&= f^{(t)}\odot c^{(t-1)} + i^{(t)} \odot g^{(t)}\label{eqn:s_lstm}\\
	h^{(t)}&= \mathrm{tanh}(c^{(t)})\odot q^{(t)}\label{eqn:h_lstm}
\end{align}
where at time $t$ the input, forget, and output gates are $i^{(t)}$, $f^{(t)}$, and $o^{(t)}$, respectively. 
The $g^{(t)}$, is the input-update value.
The memory cell state at time $t$ is $c^{(t)}$, and the output of the LSTM unit at time $t$ is $h^{(t)}$.

With the ability to handle long-term dependencies effectively, LSTMs are used for developing prediction and classification algorithms that can be used for better classifying malicious and benign PowerShell script detections. 

In the LSTM model, the number of memory units (neurons) in the LSTM layer is defined as a hyperparameter. In order to achieve this goal, LSTM units work like tiny detectives by looking for different patterns in the script. Each neuron (detective) can pay attention to different parts of the script, and they work together as a team to solve the problem. For example, one can focus on sequences like \textit{IEX} followed by a suspicious URL. Another one can look at frequent downloads associated with \textit{DownloadString} function. The other one can notice script has Base64 encoding patterns. To prevent overfitting and to make the model generalized well, the dropout phase acts as a supervisor to randomly drop some neurons during training. This process makes the model to focus on a broader range of patterns.

\subsection{Bidirectional LSTM (BiLSTM)}

The BiLSTM, an improved version of LSTM, processes data in both forward and backward directions. BiLSTM is an effective tool for tasks when the context from both directions is important, such as malicious script detection \cite{graves2005framewise, graves2013speech}. The model can capture both past and future dependencies. This feature is useful in tasks like text classification \cite{NElsayed_BiLSTM2021, boden2002guide}.

Improving the behavior of two sequence paths with BiLSTM, processing each row of PowerShell scripts using NLP, and ordering tokenized IDs can be used to calculate model layer biases and weights in a much more accurate way. 

\subsection{Transformers}

The transformer model is used to understand and generate human language \cite{vaswani2017attention, wang2019language} that is useful in applications like chatbots, translation, and text generation.

The key advantage of Transformers is self-attention \cite{vaswani2017attention}. That feature helps the model to focus on different words in a sentence at the same time. The model can understand the context rather than their sequential order. Semantic learning level is higher than other previous models. Older models read text one word at a time. However, Transformers process entire sentences at once. This makes the learning faster and more efficient \cite{manaswi2018rnn, wang2019language, vaswani2017attention, soydaner2022attention}. Transformer-based models include BERT, CodeBERT, GPT, and T5.

Transformers is a neural network architecture for sequence modeling tasks to get better predictions than RNNs and LSTMs \cite{wang2019language}. It was proposed by Vaswani et al. in 2017 \cite{vaswani2017attention} as a trainable attention mechanism for capturing complex relationships between elements of an input sequence.

It is primarily composed of an encoder-decoder structure, providing a framework for handling sequences without recurrent computations. They use positional encodings to incorporate the order of tokens into the model since they process all tokens in parallel rather than sequentially \cite{wang2019language, vaswani2017attention}.

Transformers have the Softmax function that plays an important role in normalizing the attention scores \cite{niu2021review, vaswani2017attention}.

The attention mechanism focuses on different parts of the input sequence for predictions. It calculates a weighted sum of the input tokens, where the weights are determined by the relevance of each part of the input to the current token \cite{niu2021review}.

\subsection{CodeBERT}

Our research focuses on the classification problem of malicious PowerShell scripts. In this sentimental analysis task, sequence-to-representation models are useful \cite{dos2014deep}. CodeBERT can be used \cite{feng2020codebert} as an encoder-only base transformer model to analyze scripts sentimentally on specific malicious script detection.

CodeBERT was developed based on RoBERTa \cite{liu2019roberta} and BERT \cite{devlin2018bert} models. It uses Transformer-based neural architecture and relies on a bimodal pre-trained model. It was also designed for both natural language (NL) and programming language (PL). It can acquire the semantic relationships between NL and PL. This can provide general-purpose representations that make the tasks easier, as well as natural language code search and code documentation generation. \cite{feng2020codebert}.

CodeBERT is a proven model with the effectiveness of semantic learning of code. It is a pre-trained model on both Natural Language (NL) and Programming Language (PL) that is designed for code-related tasks. CodeBERT captures semantic meaning better than LSTMs. It is also good at avoiding overfitting faster due to its pre-training. It is more suitable for script classification tasks with fine-tunable on specific tasks. Additional fine-tuning is necessary for training the model because it makes the model better for specific tasks, such as the classification of benign and malicious scripts.

The results demonstrate that CodeBERT achieves state-of-the-art performance in learning PL and NL tasks \cite{feng2020codebert}. The model is created using the multi-layer Transformer architecture \cite{vaswani2017attention, feng2020codebert}. CodeBERT is considered appropriate for fundamental classification tasks that rely on syntax analysis and code execution reports.

\section{Related Works}

Table~\ref{Comparison_of_reviews} shows a comparison of the experimental results on malicious script detections. As seen in the table, malicious script datasets are limited. Malicious script datasets range from as few as 100 samples \cite{khalid2023insight} to over 6,600 samples \cite{alahmadi2022mpsautodetect}. Model training with larger datasets is preferable for accurate performance evaluation. However, smaller datasets can limit generalization and suffer from overfitting.

Since high recall is crucial to detect as many malicious scripts as possible, the focus on recall rate is preferable rather than accuracy. Because, missing a threat is riskier than false alarms in cyber security.

Feature extraction usage is common. It is used to improve detections; however, as seen in the papers from \cite{rusak2018ast} and \cite{liguori2024power}, the models still achieve high recall scores. 

The highest recall rate (99.0\%) is shown by \cite{mimura2021static} using XGBoost with feature extraction. However, they used a relatively small dataset (480 samples), and they also reported this in the limitations.

\begin{table*}[h!]
    \centering
    \caption{Comparison of the ML results on malicious PowerShell script detections.}
    \begin{tabular}{l p{2cm} p{2cm} p{2cm}ccc}
        \hline
        \textbf{Research} & \textbf{Malicious Data} & \textbf{Feature Extr} & \textbf{Method} & \textbf{Recall \%}\\
        \hline
        \cite{rusak2018ast}&4,079&No&AST-based random forest&85.0 \\
        \cite{hendler2018detecting}&6,290&Yes&NLP-based CNN&89.0 \\
        \cite{mimura2021static}&480&Yes&XGBoost&99.0  \\
        \cite{fang2021effective}&2,000&Yes&AST-based random forest&97.6  \\
        \cite{alahmadi2022mpsautodetect}&6,609&Yes&SdAs-based XGBoost&98.0\\
        \cite{khalid2023insight}&100&Yes&Random forest classifier&87.5 \\
        \cite{liguori2024power}&1,128&No&CodeGEN + CodeGPT&85.2\\
        \hline
    \label{Comparison_of_reviews}
    \end{tabular}
\end{table*}

\section{Preparation of Experimental Analysis}

\subsection{Pipeline of ASTs} 
\label{ML_AST_Method}

AST reports include command keywords and parameters with their corresponding AST types. However, the current format is not suitable for effective data preprocessing and NLP tasks because it can reduce model training accuracy. Therefore, creating a pipeline for AST reporting is useful so that each report is clear and ready for preprocessing steps, including tokenization. In the pipeline, each command and parameter are transformed into pairs with related AST types, and the command is rewritten per row to file in JSON format.

First of all, data was labeled as malicious scripts with an integer "\textit{1}" and benign scripts with an integer "\textit{0}". After that, the pipeline processed the scripts for Abstract Syntax Trees (ASTs). The first phase converted PowerShell scripts into ASTs using PowerShell's built-in function:

\noindent
\textit{[System.Management.Automation.Language.Parser]}.

Tokenizing is a process in natural language processing (NLP) and machine learning to break down text data into word or phrase strings. Those manageable pieces are called tokens. Those meaningful elements help us to analyze, process, and feed into machine learning models \cite{kul2022detecting}.

Tokenization can be done with two different methods: character-based or word-based tokenization. We used word-based tokenization in LSTM and BiLSTM models after the process of AST representations aligning with the AST approach. The model received the words as tokens. Since this approach suffers due to out-of-vocabulary (OOV) issues, limiting OOVs is useful. 

On the other hand, Transformers has a special tokenization method that they can use subword-based tokenization such as Byte Pair Encoding (BPE), WordPiece, or SentencePiece \cite{wang2019language}.

In this study, we focus on the most frequent words in the dataset and set a limit of 6,000 unique tokens out of ~250,000 tokens. Tokens exceeding this limit are categorized as “Out of Vocabulary” (OOV) and assigned a default ID value of 1. “OOV” (Out-Of-Vocabulary) token is used for unknown words for unseen data. Both benign and malicious scripts are included in the tokenizer’s vocabulary to provide a comprehensive base for model training.

To improve the effectiveness of the tokenization process, we exluded system administration keywords. This filtering helps to avoid the addition of very common system administrator words into the vocabulary.

\subsection{Phases of Machine Learning Models} 

In order to conduct machine learning analysis, we followed the main phases. First, we import data by reading and loading the AST data. After that, we tokenize the data with padding sequences and limiting vocabulary. Dataset splitting process is defined for each model for the appropriate analysis. After fine-tuning with adjustment dropouts and hypermeter tuning, we trained the models.

Since high recall is crucial to detecting as many malicious scripts as possible, we focused on recall rate rather than accuracy because missing a threat is riskier than false alarms in cybersecurity.

\subsection{Data Splitting and Validation Method} 

We used traditional data splitting for the LSTM and BiLSTM models with 70\% training, 15\% validation, and 15\% test data as shown in table \ref{traditional_split}.

\begin{table}[h]
  \begin{center}
    \caption{LSTM and BiLSTM – Traditional Data Splitting.}
    \begin{tabular}{l|c|c|c|c}
        \hline
        \textbf{Set Total} & \textbf{Samples} & \textbf{Benign [0]} & \textbf{Malicious [1]} & \textbf{Token}\\
        \hline
        Training 70\% & 2,800 & 1,400 & 1,400 & 400  \\
        Validation 15\% & 600 & 300 & 300 & 400  \\
        Test 15\% & 600 & 300 & 300 & 400  \\
        \hline
    \end{tabular}
    \label{traditional_split}
  \end{center}
\end{table}

Also, we used K-fold cross-validation in the CodeBERT model with 80\% training and 20\% validation sets. In K-fold cross-validation, the data is divided into K parts, and each fold validates on a different 20\% of the data while training on the remaining 80\%. Therefore, there is no need to specify a separate testing set. Because each fold acts as both training and validation sets as seen in Table \ref{codebert_split}.

\begin{table}[h]
  \begin{center}
    \caption{CodeBERT K-5 Cross-Validation}
    \begin{tabular}{l|c|c|c|c}
        \hline
        \textbf{Set Total} & \textbf{Samples} & \textbf{Benign [0]} & \textbf{Malicious [1]} & \textbf{Token}\\
        \hline
        Training 80\% & 3,200 & 1,600 & 1,600 & 400  \\
        Validation 20\% & 800 & 870 & 300 & 400  \\
        \hline
    \end{tabular}
    \label{codebert_split}
  \end{center}
\end{table}

\subsection{Model Definition - LSTM and BiLSTM models}

The LSTM (Long Short-Term Memory) and BiLSTM (Bidirectional LSTM) models are proposed with a custom-built model using the embedding layer, LSTM, dropout, and dense layers.

The embedding layer dimension is 128, and the tokens are represented with 128 vector dimensions. For example, if a contains the word “\textit{DownloadString}”, it is first converted into a numeric token such as “295” as if it is the 245th word (token) in the vocabulary. After that, it is transformed into a 128-dimensional vector with learned weights such as [0.12,-0.56,0.89,...,0.34](128 values); note that the vocabulary size is 6,000 tokens. The number of trainable word embeddings is 768,000 (6,000 × 128). The sequence length is 400 to ensure all input sequences have the same length, avoiding exceeding padding. 

In the hidden LSTM layer, the number of units is 64 with a balanced learning capacity and expected computational efficiency. Since the problem is classification of malicious an benign scripts, the return sequences setting is defined as false. This gives outputs only the last hidden state focusing on sequence-level understanding. 

The dropout layer is set up with 50\% dropout rate to reduce overfitting by randomly disabling 50\% of neurons during training. This step helps for generalizated model creation. The activation is ReLU in this layer that it provides LSTM’s output into a higher-dimensional space before final classification. Also, we used a second drpout layer to prevent overfitting before final output. 

This approach is also applied to the BiLSTM model, that it uses two-direction learning compared with LSTM learning. That makes BiLSTM a heavier model training process than LSTM. The adoption of bidirectional LSTM models offers distinct advantages over conventional LSTM models \cite{lample2016neural, do2019hate, hochreiter1997long}. The BiLSTM architecture processes the sequence in two directions, allowing the model to capture both past and future dependencies \cite{graves2013speech}. This improves performance on tasks where context matters in both aspects. This feature is useful in tasks such as sequence classification, named entity recognition, and machine translation \cite{NElsayed_BiLSTM2021}. 

\subsection{Model Definition - CodeBERT}

As explained in the background section, CodeBERT is a sequence-to-representation transformer model designed for a variety of text-based tasks, including the classification of labeled texts for model training \cite{feng2020codebert}.

CodeBERT model is proposed as a fine-tuned pre-trained model. Each parameter in the CodeBERT model is carefully chosen to build a better model for high recall rate. 

First, the model uses RoBERTa-based tokenization based on Hugging Face model build. It breaks words into sub-word units for better representation. Also, it uses "Out-of-Vocabulary (OOV)" words by splitting rare or unknown words into smaller components.

The learning rate is "0.00002" because to prevent sudden drastic changes during training. As a real world example, small steps are safer while learning walking. Therefore, this low learning rate helps the model to avoid dramatic forgetting.

The batch size is 8 to process a small number of scripts at once. In each iteration, the checking process on the small number of scripts provides memory efficiency with limited resources.

The weight decay value is 0.01 to avoid too much memorizing everyting and to much attention on a pattern that can cause overfitting. This process adds a penalty for large weights ensuring generalized model creation.  
 
The model uses AdamW (Adaptive Moment Estimation with Weight Decay) optimizer to clean up learning to focus on the important patterns in classification. This process also helps us prevent overfitting.
 
As a loss function, the model uses "Cross-Entropy Loss" that is a typical function for classification problems. This helps the model to learn probabilistic outputs. As a real world example, this function acts as a supervisor and uses a scoreboard to check the validations.
 
We use K-Fold Cross-Validation (K=5) for model training accuracy validation to monitor overfitting. The model is tested across multiple splits of data. This process reduces dependency on a single train-test split. It can detect overfitting and assess generalization.
 
The model has 12 transformer layers and 12 attention heads for better self-attention mechanisms. Each layer processes hierarchical representations, checking complex patterns in the malicious and benign scripts. The model can focus on different parts simultaneously with the attention heads.
 
In this model, the hidden dimensionality size is 768. Each token is represented as a 768-dimensional vector in transformer layers.  
 
Unlike the use of tokenizers in LSTM-based models, the pre-trained CodeBERT model is applied to modified AST reports using its pre-trained transformer tokenizer. Padding for each array is kept to the maximum length of a specific AST command block, and truncation is applied as needed \cite{feng2020codebert}.

As noted in the literature review on CodeBERT \cite{feng2020codebert}, the main training application, using five main code syntaxes and functionalities, improves the efficiency of the built-in dataset by accommodating different syntax variations in scripts.

Therefore, fine-tuning the pre-trained CodeBERT model on PowerShell-based AST reports mitigates the complexity arising from different code representations within the dataset while also improving the accuracy of script classification.

\begin{figure}
    \centering
    \includegraphics[width = 11 cm, height= 5 cm, alt={The diagram blocks of BiLSTM and the information flow between blocks.}]{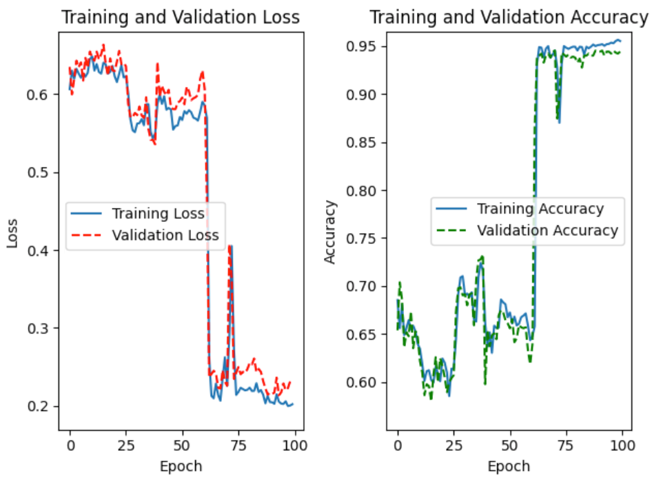}
    \caption{BiLSTM model training and validation loss and accuracy results.}
    \label{bilstm_results}
\end{figure}

\section{Experimental Results}

\begin{table*}[ht!]
    \centering
    \caption{Performance Metrics for Related Works and Our Examination.}
    \begin{tabular}{l|c|c|c|c}
        \hline
        \textbf{Model} & \textbf{Acc. \%} & \textbf{Precision \%} & \textbf{Recall \%} & \textbf{F1-Score} \\
        \hline
        Non-AST LSTM & 90.1 & 94.5 & 93.5 & 91.8 \\
        AST-based LSTM & 93.3 & 98.5 & 95.3 & 93.8 \\
        AST-based BiLSTM & 95.3 & 92.6 & 91.8 & 92.2 \\
        Non-AST CodeBERT & 96.2 & 96.1 & 93.6 & 96.3 \\
        AST-based CodeBERT & 96.2 & 96.1 & \textbf{96.6} & 96.3 \\
        \hline 
    \end{tabular}
    \label{combined_models_evaluation}
\end{table*}

The models were evaluated using the fundamental metrics of accuracy, precision, recall, and F1 score. However, we focused on recall scores. Recall assesses the model's ability to capture all relevant items within the dataset to measure how successful the model is against false negatives that can be critical errors in cybersecurity.

The models were evaluated using fundamental metrics: accuracy, precision, recall, and F1 score. However, we focused on the recall score tto measure how well the model performs against false negatives. Detecting malicious scripts as benign scripts is a critical error in cybersecurity domain that it might allow threat actors access the networks.

The BiLSTM model was trained for over 100 epochs to monitor the overfitting. The overfitting signals are observed at the end of the $\mathrm{15^{th}}$ epoch, and early stopping was applied at the $\mathrm{16^{th}}$ epoch. The model was saved in the best result with the feature of early stopping. The training and validation loss and accuracy scores is shown in Fig. \ref{bilstm_results}.

Unlike BiLSTM, in the training of the conventional LSTM model, early stopping reached the $\mathrm{71^{th}}$ epoch of the training process for the expected range of validation loss and validation accuracy. LSTM's best result was also saved with the feature of early stopping.

The LSTM, BiLSM, and CodeBERT models were trained with AST-based tokenization and Non-AST tokenization to check the effectiveness of the AST-based approach as well.

Table \ref{combined_models_evaluation} shows the performance metrics of three different models with the AST-based approach options.

\section{Key Findings}

The results indicates that AST-based fine-tuned CodeBERT achieved a high recall rate of 96.6\% proving the usage of AST can improve the recall score of CodeBERT model. 

Also, the results showed that AST-based fine-tuned CodeBERT can be used for an effective model with the proposed fine-tuning method without feature extraction. This can help us reduce computational overhead. 

The proposed model provides a better solution compared the other solutions in the literature even though there are other higher recall rates. Because AST-based fine-tuned CodeBERT achieved a high recall rate of 96.6\% without feature extraction and using a balanced dataset with the pre-trained model.

Although our model does not achieve top results compared to other models, it improves performance in terms of efficiency. Also, while setting up fine-tuned model, we used optimized hyperparameters especially for the model layer dimensions to avoid higher computational need. 

Furthermore, our approach does not include obfuscation and de-obfuscation processes that can cause more processing steps by bringing more computational overhead. The current models received the tokens as is in terms of complexity and entropy. This approach also improved efficiency while having a high recall rate.

\section{Conclusion}

This study conducted an experimental study on detecting PowerShell-based fileless cryptojacking scripts. The experimental results showed that AST-based fine-tuned CodeBERT achieved a high recall rate of 96.6\%. This result proved the importance of using AST integration. The model achieves this high recall score without feature extraction. This approach helped us to reduce computational overhead while maintaining effectiveness. 

Although some other models in the literature provided higher recall rates, our approach brought a more efficient solution. This was achieved with a balanced dataset and an AST-based fine-tuned pre-trained model. The pre-trained CodeBERT model was optimized with appropriate hyperparameters, minimizing computational needs. Additionally, our method avoided obfuscation and de-obfuscation processes. This approach reduced complexity while maintaining a high recall rate.

As a limitation of this study, dataset creation is extremely challenging due to the difficulty of accessing malicious scripts. This can be solved by joint collaboration between academic researchers and security engineers in cybersecurity services.

\section{Acknowledgements}
We would like to thank Talha Aydin, a machine learning researcher in the Department of Computer Science and Information Systems at De Anza College, California, for providing valuable feedback.


\section{Submission Declaration}
\begin{itemize}
    \item The work described has not been published previously except in the form of a preprint, and academic thesis.
    \item The article's publication is approved by all authors and tacitly or explicitly by the responsible authorities where the work was carried out.
\end{itemize}

\section{Authorship}
All authors have made substantial contributions including conception and design of the study, drafting the article and perform fully revision of the content. 

\section{CRediT Roles}
Said Varlioglu: Conceptualization, Conceptualization, Investigation, Formal analysis, Visualization, Validation, Writing – original draft. Nelly Elsayed: Conceptualization, Methodology, Supervision, Writing – original draft, Writing – review \& editing. Murat Ozer: Validation, Data curation, Writing – review \& editing. Zag ElSayed: Validation, Formal analysis, Writing – review \& editing. John M. Emmert: Supervision, Project administration.

%\section{Declaration of Interests}
%The authors declare that they have no known competing financial interests or personal relationships that could have appeared to influence the work reported in this paper.

\section{Funding}
This research did not receive any specific grant from funding agencies in the public, commercial, or not-for-profit sectors.

\section{Disclosure statement}
No potential competing interest was reported by the author(s).

\section{Declaration of Generative AI Use}
During the preparation of this work the author(s) used Grammarly in order to assist in spell-check and grammar-check. The author reviewed, validated, and edited all content to ensure accuracy, integrity, and alignment with scholarly standards and take(s) full responsibility for the content of the published article.

%During the preparation of this work the author(s) used ChatGPT, Gemmini, Scholar Lab, and Grammarly in order to assist in draft refining, and organizing sections of the manuscript, spell-check and grammar-check, and Latex syntax revision. The author reviewed, validated, and edited all content to ensure accuracy, integrity, and alignment with scholarly standards. No AI tools were used for data analysis, interpretation of findings, or generation of scientific claims. After using this tool/service, the author(s) reviewed and edited the content as needed and take(s) full responsibility for the content of the published article.

\section{Data availability statementt}
Due to the sensitive nature of the collected data that contains malicious PowerShell scripts, the data would be shared only based on request and providing signing a responsibility usage form by the requester that includes the purpose of use and plan. After reviewing and proceed thought the office of research the access may or may not be provided depending on the office of research decision.
%This study did not generate or analyze any datasets. All findings are based on conceptual analysis, publicly available literature, and previously published sources. Therefore, no datasets were created, used, or are available for sharing.

\section{Biographical note}
\noindent \textit{Said Varlioglu (Ph.D.)}, Digital Forensics and Incidence Response Engineer. Recieved his Ph.D. at Information Technology from Univeristy of Cincinnati in 2025.\\
\textit{Nelly Elsayed (Ph.D.)}, Associate Professor, Leader and Founder of the Applied Machine Learning and Intelligence (AMLI) Lab at the University of Cincinnati. \\
\textit{Murat Ozer (Ph.D.)}, Associate Professor at the University of Cincinnati.\\
\textit{Zag ELSayed (Ph.D.)}, Assistant Professor at the University of Cincinnati and a Fellow Researcher at the Cincinnati CHildren's Hospital Medical Center\\
\textit{John Emmert (Ph.D.)}, Professor in the Department of Electrical Engineering and Computer Science and Director of the NSF Center for Hardware and Embedded Systems Security and Trust (CHEST) IUCRC. 
% \bibliographystyle{elsarticle-num} 
%\bibliography{main}



\end{document}